\documentclass[epjCONF,columns]{svjour} 
\usepackage{graphics}
\usepackage[varg]{txfonts}
\usepackage[latin1]{inputenc}
\session-title{2011 Hadron Collider Physics symposium (HCP-2011)}
\newcommand{\pp}{$p$-$p$ }
\newcommand{\Zmumu}{$Z\rightarrow\mu\mu$ }
\newcommand{\JPsimumu}{$J/\psi\rightarrow\mu\mu$ }
\newcommand{\pT}{$p_{T}$ }
\newcommand{\Eta}{$|\eta|$}
\newcommand{\les}{$<$}
\newcommand{\fb}{fb$^{-1}$}
\newcommand{\Ecm}{$\sqrt{s}=7$ TeV}
\begin{document}
\title{Measurement of muon momentum resolution of the ATLAS detector}
\author{Antonio Salvucci\inst{1}\fnmsep\thanks{\email{a.salvucci@science.ru.nl}} on behalf of the ATLAS Collaboration}
\institute{Radboud University Nijmegen and Nikhef}
\abstract{
The ATLAS detector has been designed to have good muon momentum resolution up 
to momenta in the TeV range. The muon momentum resolution of the ATLAS spectrometer
has been measured with \pp collision data recorded in 2011. The measurement 
combines the di-muon mass resolution in \JPsimumu and \Zmumu decays with measurements 
of the alignment accuracy of the detector based on straight muon tracks, which are 
acquired with special runs without magnetic field in the ATLAS detector.
}
\maketitle
\section{Introduction}
\label{intro}
The physics programme of the ATLAS Experiment \cite{ATLAS} at the LHC includes investigations
of many processes with final state muons. The ATLAS detector is equipped with a Muon Spectrometer
(MS) optimized to provide a momentum measurement with a relative resolution better than 3\% over 
a wide \pT range and 10\% at \pT=1 TeV, where \pT is the muon momentum component in the plane 
transverse to the beam axis. The momentum in the MS is measured from the deflection of the muon 
trajectory in the magnetic field generated by a system of air-core toroid coils. The MS 
track is reconstructed using three layers of precision drift tube (MDT) chambers in the 
pseudorapidity\footnote{The pseudorapidity is $\eta=-\ln(\tan(\theta/2))$, where $\theta$ is the 
polar angle with respect to the beam line.} range \Eta\les2, and two layers of MDT chambers behind 
one layer of cathode strip chambers (CSC) for 2$\leq$\Eta\les2.7.

An additional determination of the muon momentum is provided by the Inner Detector (ID) for 
\Eta\les2.5. The ID is composed of three detectors providing coordinate measurements for track 
reconstruction inside a solenoidal magnetic field of 2 T. Close to the interaction point 
it features a silicon pixel detector, surrounded by a silicon strip detector (SCT). In the outermost
part there is a transition radiation straw tube tracker (TRT) with a coverage of \Eta\les1.9.

Muons entering in this analysis are reconstructed as \emph{combined muons}. The underlying muon 
identification is described in \cite{MuonId} and relies on the principle that first separate 
tracks are measured in ID and MS before they are reconstructed as a single trajectory. This yields
a higher momentum resolution than could be achieved using the individual tracks.

This paper documents the muon momentum resolution in the first pass of reconstruction of \pp 
collision data collected in 2011 corresponding to an integrated luminosity of 2.54 \fb. The first
pass of reconstruction uses preliminary calibration and alignment.

\section{Parametrization of the momentum resolution as a function of \pT and $\eta$}
\label{sec:param}
The relative momentum resolution, $\sigma(p)/p$, originates from different effects \cite{MS,ID}.
The ATLAS MS is designed to provide a momentum resolution as a function of the $\eta$ and $\phi$.
For a given value of $\eta$, the resolution can be parametrized as a function of \pT:
\begin{equation}
  \frac{\sigma(p)}{p}=\frac{p_{0}^{MS}}{p_{T}}\oplus p_{1}^{MS}\oplus p_{2}^{MS}\cdot p_{T}
\end{equation}
where $p_{0}^{MS}$, $p_{1}^{MS}$, $p_{2}^{MS}$ are coefficients related to the energy loss in the 
calorimeter material, multiple scattering and intrinsic resolution terms, respectively.

For the ID, the curvature measurement depends on the track length of the muon in the active material,
which is reduced close to the edge of the TRT fiducial volume. This results in a uniform response in the 
central part and a rapid worsening beyond this region. The approximate parametrization of the 
resolution is
\begin{eqnarray}
  \frac{\sigma(p)}{p}&=&p_{1}^{ID}\oplus p_{2}^{ID}\cdot p_{T}\;\;\;\;\;\;\;\;\;\;\;\;\;\;\;\;\;\;\mbox{for \Eta\les1.9}\\
  \frac{\sigma(p)}{p}&=&p_{1}^{ID}\oplus p_{2}^{ID}\cdot p_{T}\frac{1}{\tan^{2}(\theta)}\;\;\;\;\;\;\;\mbox{for \Eta$>$1.9}
\end{eqnarray}
where $p_{1}^{ID}$, $p_{2}^{ID}$ are the multiple scattering and the intrinsic resolution terms, respectively.

In this study, four regions in pseudorapidity are distinguished:
\begin{itemize}
\item[-] \textit{Barrel}: covering 0\les\Eta\les1.05;
\item[-] \textit{Transition}: covering 1.05\les\Eta\les1.7;
\item[-] \textit{End-caps}: covering 1.7\les\Eta\les2.0;
\item[-] \textit{CSC/No-TRT}: covering 2.0\les\Eta\les2.5.
\end{itemize}
These four regions are studied using \Zmumu decays.

\section{Combined fit to the muon resolution components}
\label{sec:fit}
The process \Zmumu is used to study the momentum resolution and to determine the corrections 
needed for simulation in each $\eta$ region of the detector. We use two quantities:
\begin{itemize}
\item[i)] the width of the reconstructed di-muon invariant mass peak at the Z pole, which is 
  a convolution of the natural width of the Z boson and the muon momentum resolution;
\item[ii)] the difference between the independent momentum measurements of the ID and MS for 
  combined muons, which is sensitive to the quadratic sum of the ID and MS momentum resolutions.
  This difference is weighted by the muon electric charge ($q/p_{T}^{ID}-q/p_{T}^{MS}$): this 
  disentangles systematic effects of the curvature due to local misalignements from the overall 
  intrinsic resolution, reducing the bias on the estimation of the resolution and correction 
  parameters.
\end{itemize}

\subsection{Global fit procedure}
Using the previous inputs, the measurements of the MS and ID momentum resolution are obtained
using a Monte Carlo template technique, based on a ``global'' fit procedure with:
\begin{itemize}
  \item[i)] \textit{a template fit to reconstructed Z lineshape}:\\
    we allow for momentum resolution smearing in the fit to the Z lineshape obtained 
    from MS and ID tracks and a combined fit with events with muons in different detector 
    regions. In this case we are sensitive to $\sigma_{mult.scatt.}\oplus\sigma_{intrinsic}$
  \item[ii)] the above, plus \textit{template fit to ($q/p_{T}^{ID}-q/p_{T}^{MS}$) distribution}:
    we allow for momentum resolution smearing in the fit, that compute several bins of 
    \pT, keeping regions separated. This case is sensitive to $\sigma_{ID}\oplus\sigma_{MS}$
  \item[iii)] \textit{external constraints on MS alignment and multiple scattering in ID and MS}
    (see \ref{sec:constraints})
\end{itemize}
Concerning the smearing, for both the MS and the ID, the transformation of \pT is
\begin{equation}
  p'_{T}=p_{T}\left(1+g\Delta p_{1}^{ID,MS}+g\Delta p_{2}^{ID,MS} p_{T}\right)
\end{equation}
where $p'_{T}$ indicates the simulated muon \pT after applying the correction $\Delta p_{i}^{ID,MS}$, 
while $g$ is a normally distributed random number with mean 0 and width 1.

\subsection{External constraints to the combined fit}
\label{sec:constraints}
In the fitting procedure, additional knowledge is introduced from independent studies, 
both for the ID and the MS. This reduces the correlation among the multiple scattering 
and the detector resolution terms in the fit, resulting in smaller uncertainties on the 
fitted parameters.

For the ID, the correction to the multiple scattering term in the ID, $\Delta p_{1}^{ID}$, 
is fixed to a value of zero, due to a knowledge of the ID material of 0.05\%. 

For the MS, the multiple scattering term, $\Delta p_{1}^{MS}$, is a free parameter of the fit. 
The energy loss of muons is mainly concentrated in the calorimeter and has been measured in 
\cite{eloss}. Its contribution to the overall MS resolution in the \pT range from 20 GeV to 
100 GeV is negligible and so no additional contribution for the energy loss, $\Delta p_{0}^{MS}$, 
is included. For the intrinsic resolution term, $\Delta p_{2}^{MS}$, the best estimate of the 
alignment accuracy is applied. This is the result of studies from samples of straight tracks 
obtained in periods of collision data taken with no magnetic field in the muon system. The 
constraints on $\Delta p_{2}^{MS}$ are reported on Table \ref{tab:p2MS_const}.
\begin{table}
  \caption{Constraints on the intrinsic resolution term}
  \label{tab:p2MS_const}
  \centering
  \begin{tabular}{ccc}
    \hline\noalign{\smallskip}
    $\eta$ region & Constraint on $\Delta p_{2}^{MS}$ (TeV$^{-1}$)  \\
    \noalign{\smallskip}\hline\noalign{\smallskip}
    barrel     & 0.143 $\pm$ 0.030 \\
    transition & 0.312 $\pm$ 0.050 \\
    end-caps   & 0.200 $\pm$ 0.050 \\
    CSC/No-TRT & 0.408 $\pm$ 0.050 \\
    \noalign{\smallskip}\hline
  \end{tabular}
\end{table}

\section{Combined fit results}
\label{sec:results}
The constraints on the $\Delta p_{i}$ parameters are applied in the combined fit by adding a penalty term
$\sum_{i}\left(\frac{\Delta p_{i}-a_{i}}{\sigma_{a_{i}}}\right)^{2}$ to the total $\chi^{2}$ being minimized,
where $a_{i}$ is the expectation value and $\sigma_{a_{i}}$ the associated uncertainty for each of the 
constrained $\Delta p_{i}$ parameters. The fitted corrections parameters are provided in 
Table \ref{tab:corrections} together with their statistical and systematic uncertainties.
\begin{table}[h]
  \caption{Set of corrections to be applied on the \pT parametrization of the simulated resolution 
    in the MS and ID to reproduce the one in data. Systematic errors on the transition region are 
    due to a different definition of the region itself (1.2\les\Eta\les1.7)}
  \label{tab:corrections}
  \centering
  \begin{tabular}{ccc}
    \hline\noalign{\smallskip}
    $\eta$ region & $\Delta p_{1}^{MS}$ (\%) & $\Delta p_{2}^{MS}$ (TeV$^{-1}$) \\
    \noalign{\smallskip}\hline\noalign{\smallskip}
    barrel     & 1.80 $\pm$ 0.05            & 0.095 $\pm$ 0.016 \\
    transition & 3.17 $\pm$ 0.15 $\pm$ 0.22 & 0.250 $\pm$ 0.026 $\pm$ 0.067 \\
    end-caps   & 1.23 $\pm$ 0.11            & 0.169 $\pm$ 0.069 \\
    CSC/No-TRT & 0.52 $\pm$ 0.58            & 0.453 $\pm$ 0.028 \\
    \noalign{\smallskip}\hline\noalign{\smallskip}
    $\eta$ region & $\Delta p_{1}^{ID}$ (\%) & $\Delta p_{2}^{ID}$ (TeV$^{-1}$) \\
    \noalign{\smallskip}\hline\noalign{\smallskip}
    barrel     & 0 & 0.283 $\pm$ 0.011 \\
    transition & 0 & 0.736 $\pm$ 0.022 $\pm$ 0.567 \\
    end-caps   & 0 & 0.871 $\pm$ 0.017 \\
    CSC/No-TRT & 0 & 0.050 $\pm$ 0.001 \\
    \noalign{\smallskip}\hline
  \end{tabular}
\end{table}
\begin{figure}[h]
  \centering
  \resizebox{0.495\columnwidth}{!}{\includegraphics{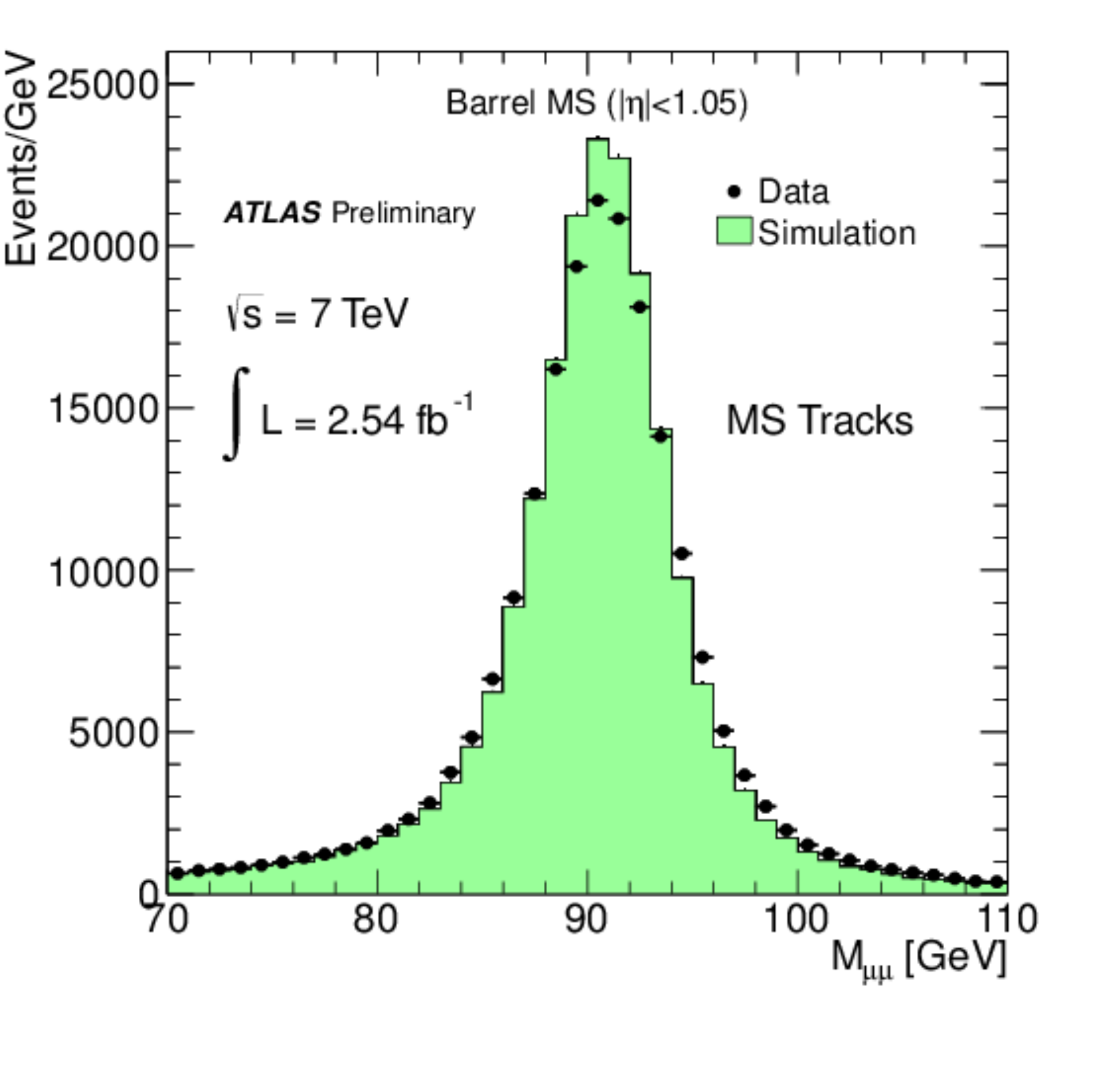}}
  \resizebox{0.495\columnwidth}{!}{\includegraphics{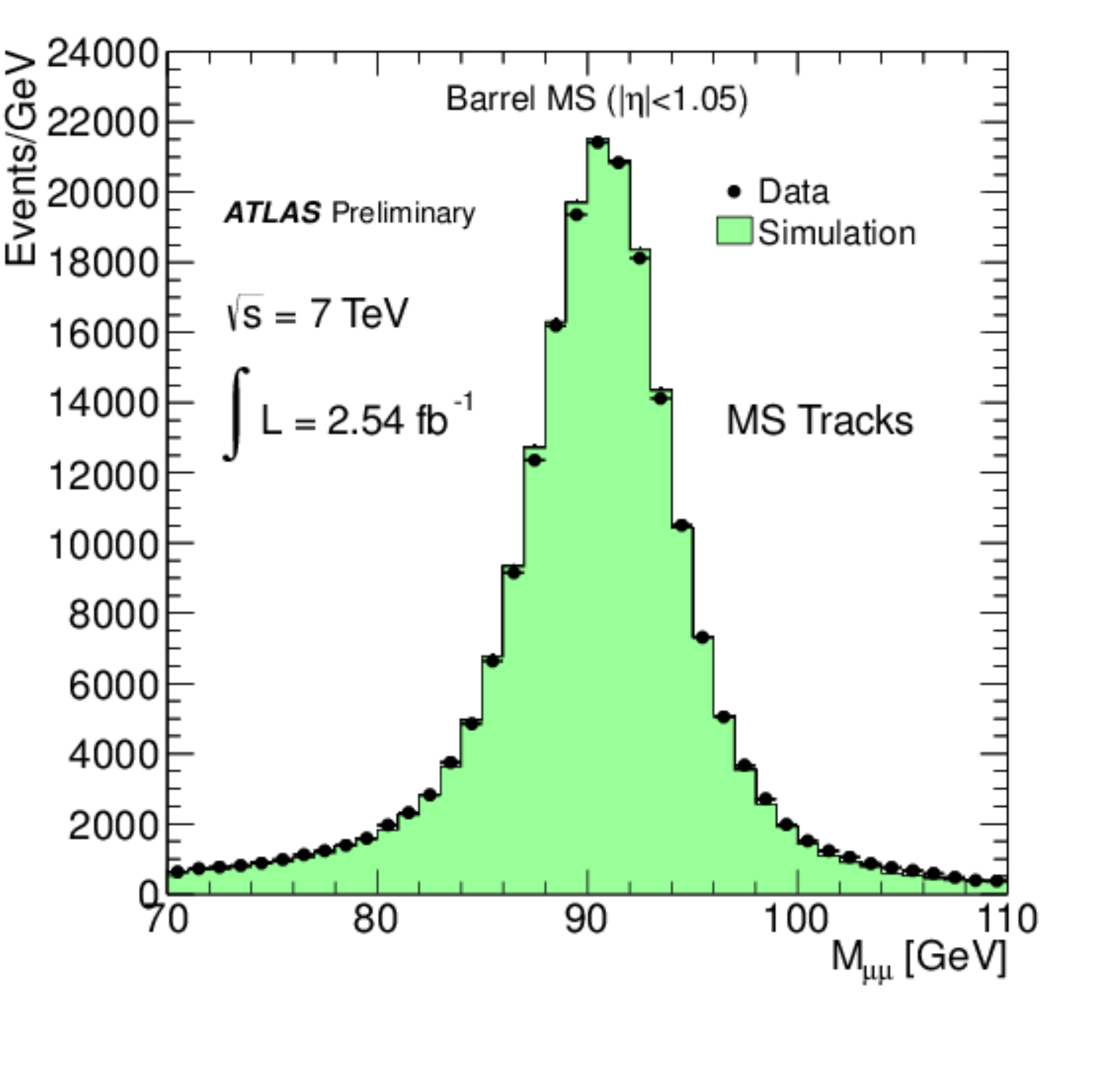}}
\caption{Examples of Z lineshape fit for MS tracks in the barrel region, 
  before (on the left) and after fit (on the right).}
\label{fig:fit_example}
\end{figure}
Figure \ref{fig:fit_example} shows an example of Z lineshape fit for MS tracks in 
the barrel region, before (left) and after (right) the fit.

The values of the correction parameters quantify the increase in momentum resolution in data
when compared to simulation. The full parametrization of the experimental momentum resolution
is obtained by quadratically adding the uncorrected simulated resolution terms of equations 
1-3 and the corresponding parameters from Table \ref{tab:corrections}. 
The results for the full parametrization are listed in Table \ref{tab:resovalues}.
\begin{table}
  \caption{Resolution parametrization as defined in equations 1-3 in the MS and ID.}
  \label{tab:resovalues}
  \centering
  \begin{tabular}{cccc}
    \hline\noalign{\smallskip}
    $\eta$ region & $p_{0}^{MS}$ (TeV) & $p_{1}^{MS}$ (\%) & $p_{2}^{MS}$ (TeV$^{-1}$) \\
    \noalign{\smallskip}\hline\noalign{\smallskip}
    barrel     & 0.25 $\pm$ 0.01 & 3.27 $\pm$ 0.05 & 0.168 $\pm$ 0.016 \\
    transition & 0               & 6.49 $\pm$ 0.26 & 0.336 $\pm$ 0.072 \\
    end-caps   & 0               & 3.79 $\pm$ 0.11 & 0.196 $\pm$ 0.069 \\
    CSC/No-TRT & 0.15 $\pm$ 0.01 & 2.82 $\pm$ 0.58 & 0.469 $\pm$ 0.028 \\
    \noalign{\smallskip}\hline\noalign{\smallskip}
    $\eta$ region & $p_{0}^{ID}$ (TeV) & $p_{1}^{ID}$ (\%) & $p_{2}^{ID}$ (TeV$^{-1}$) \\
    \noalign{\smallskip}\hline\noalign{\smallskip}
    barrel     & n.a & 1.55 $\pm$ 0.01 & 0.417 $\pm$ 0.011 \\
    transition & n.a & 2.55 $\pm$ 0.01 & 0.801 $\pm$ 0.567 \\
    end-caps   & n.a & 3.32 $\pm$ 0.02 & 0.985 $\pm$ 0.019 \\
    CSC/No-TRT & n.a & 4.86 $\pm$ 0.22 & 0.069 $\pm$ 0.003 \\
    \noalign{\smallskip}\hline
  \end{tabular}
\end{table}

\section{Measured resolutions as a function of \pT}
\label{sec:resoplots}
\begin{figure}[h]
  \centering
  \resizebox{0.675\columnwidth}{!}{\includegraphics{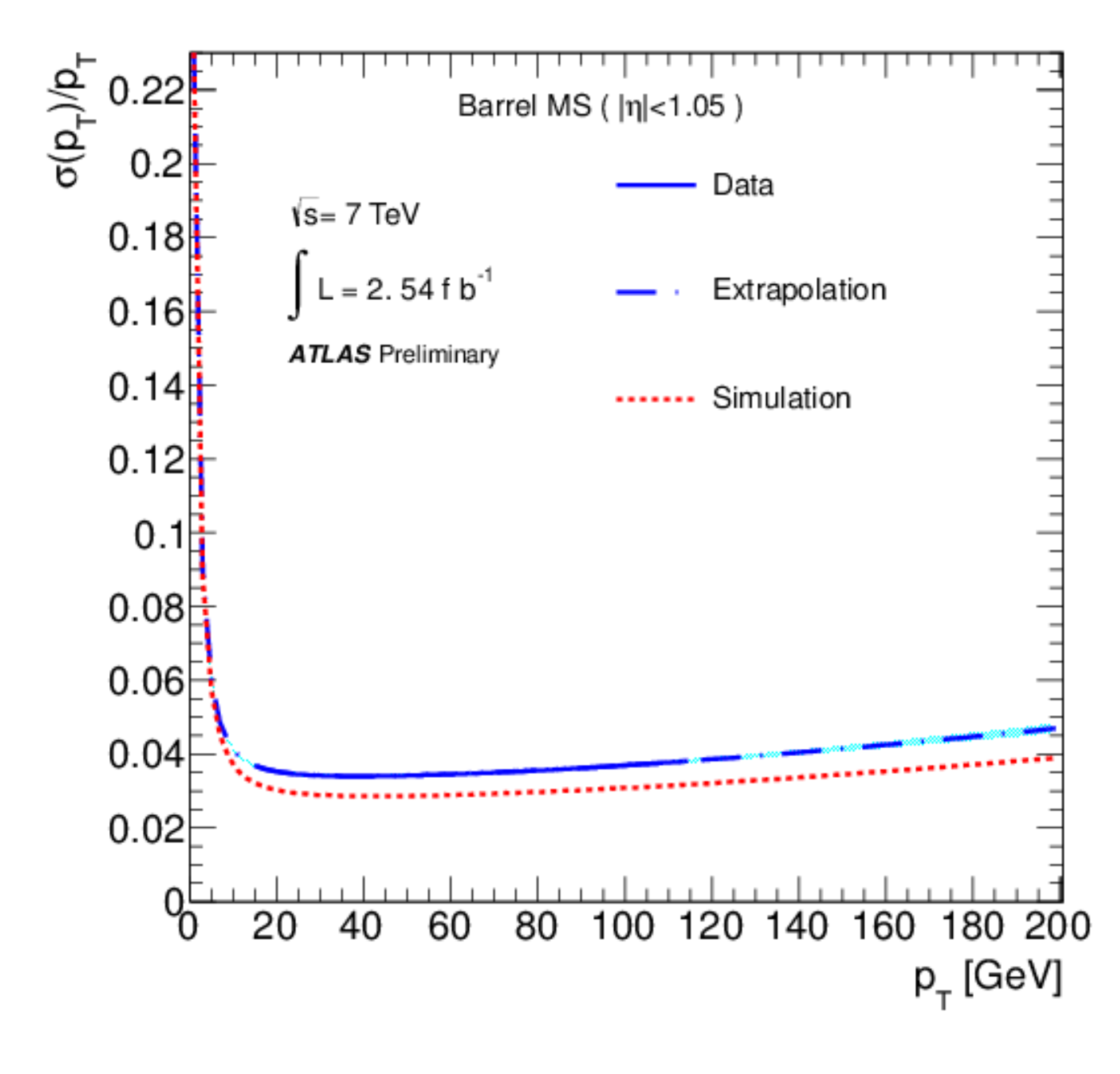}}
  \resizebox{0.675\columnwidth}{!}{\includegraphics{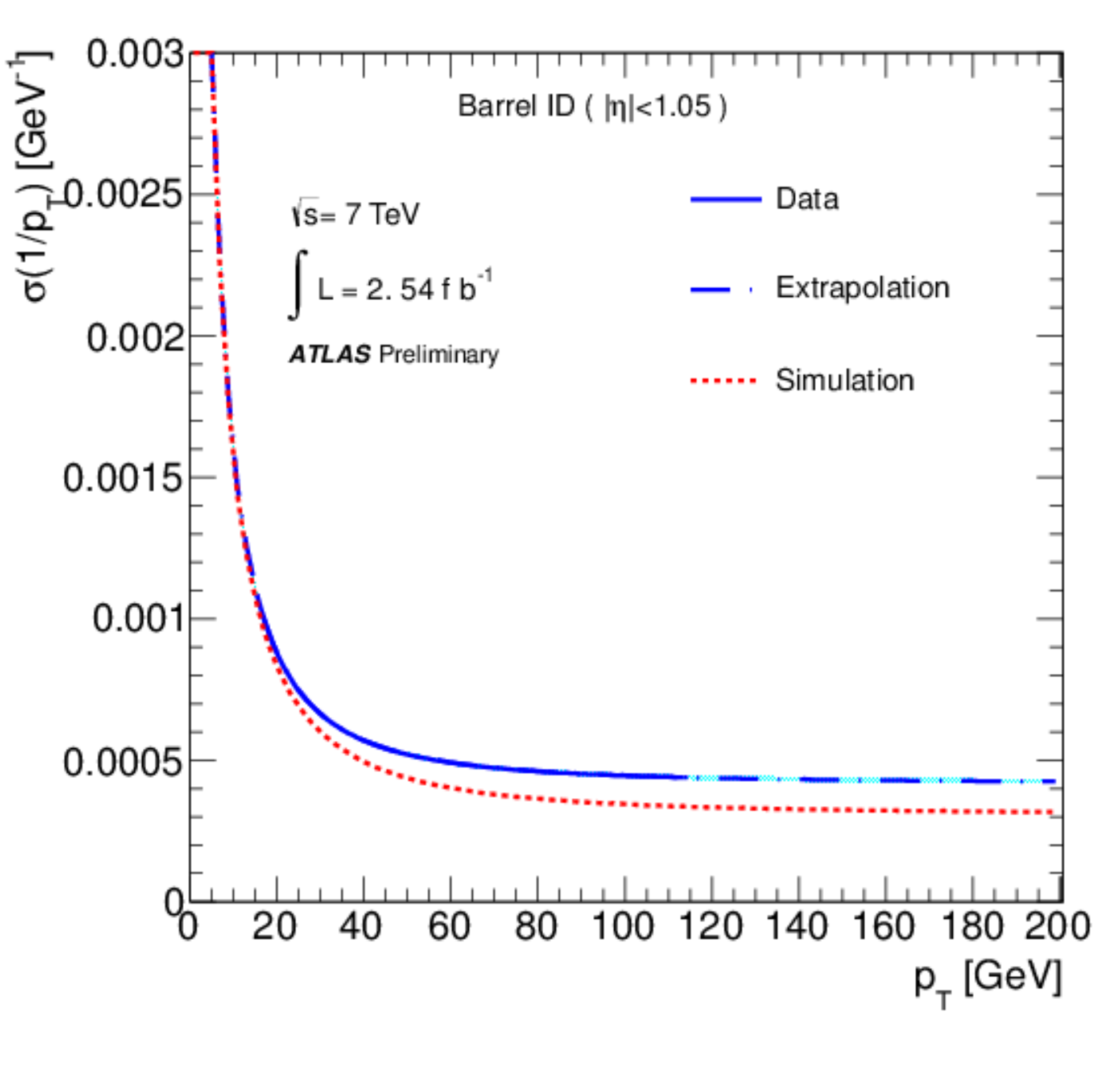}}
  \caption{Resolution curve from the fitted parameter values on the MS (top) and the ID (bottom)
    in collision data and simulation as a function of the muon \pT, for the barrel region. The 
    solid blue line shows determinations based on data, the dashed blue line shows the extrapolation 
    to \pT range not accessible in this analysis and the dashed red line shows the determinations 
    from simulation.}
  \label{fig:plotMSID}
\end{figure}

\begin{figure}
  \centering
  \resizebox{0.495\columnwidth}{!}{\includegraphics{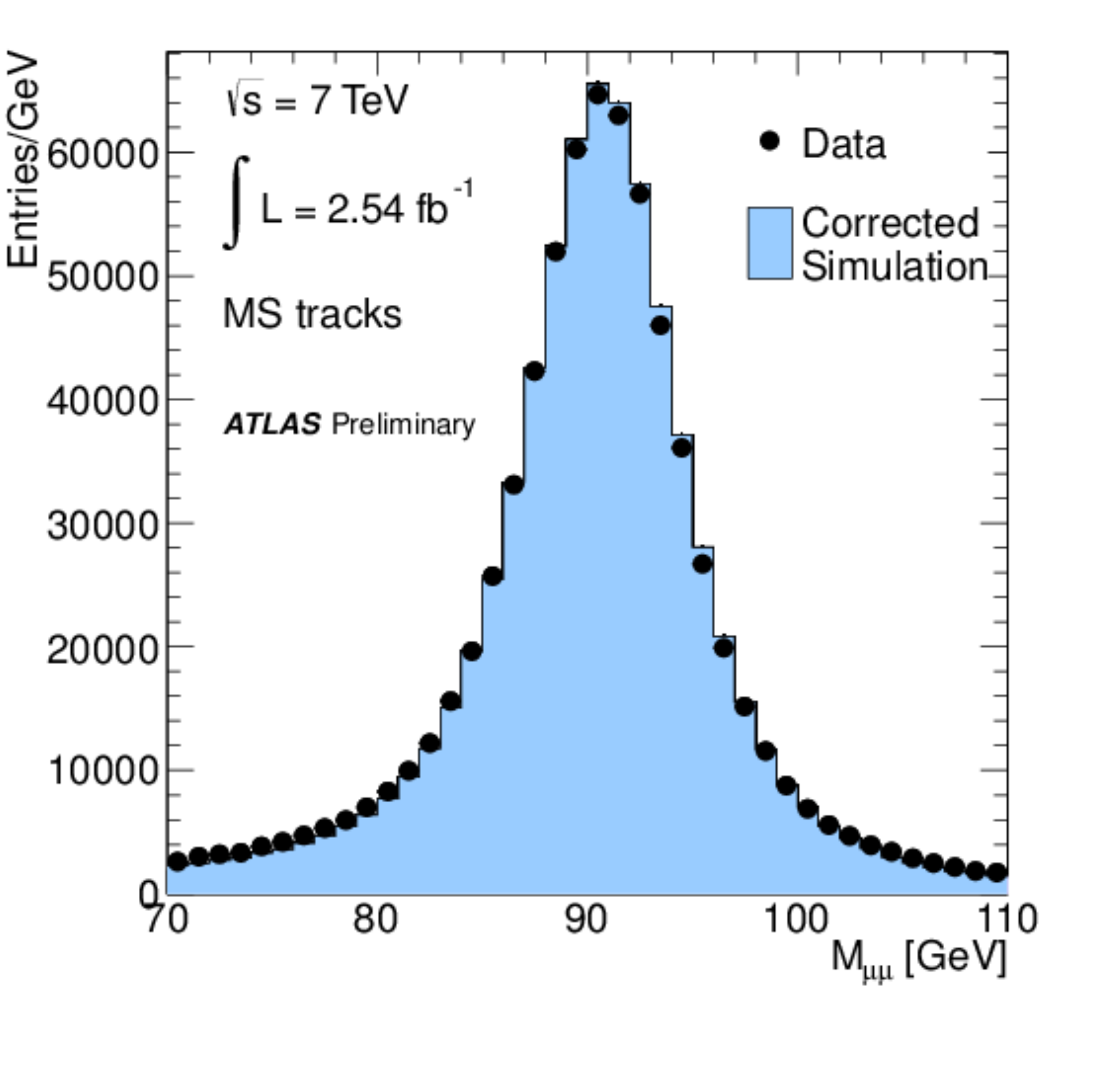}}
  \resizebox{0.495\columnwidth}{!}{\includegraphics{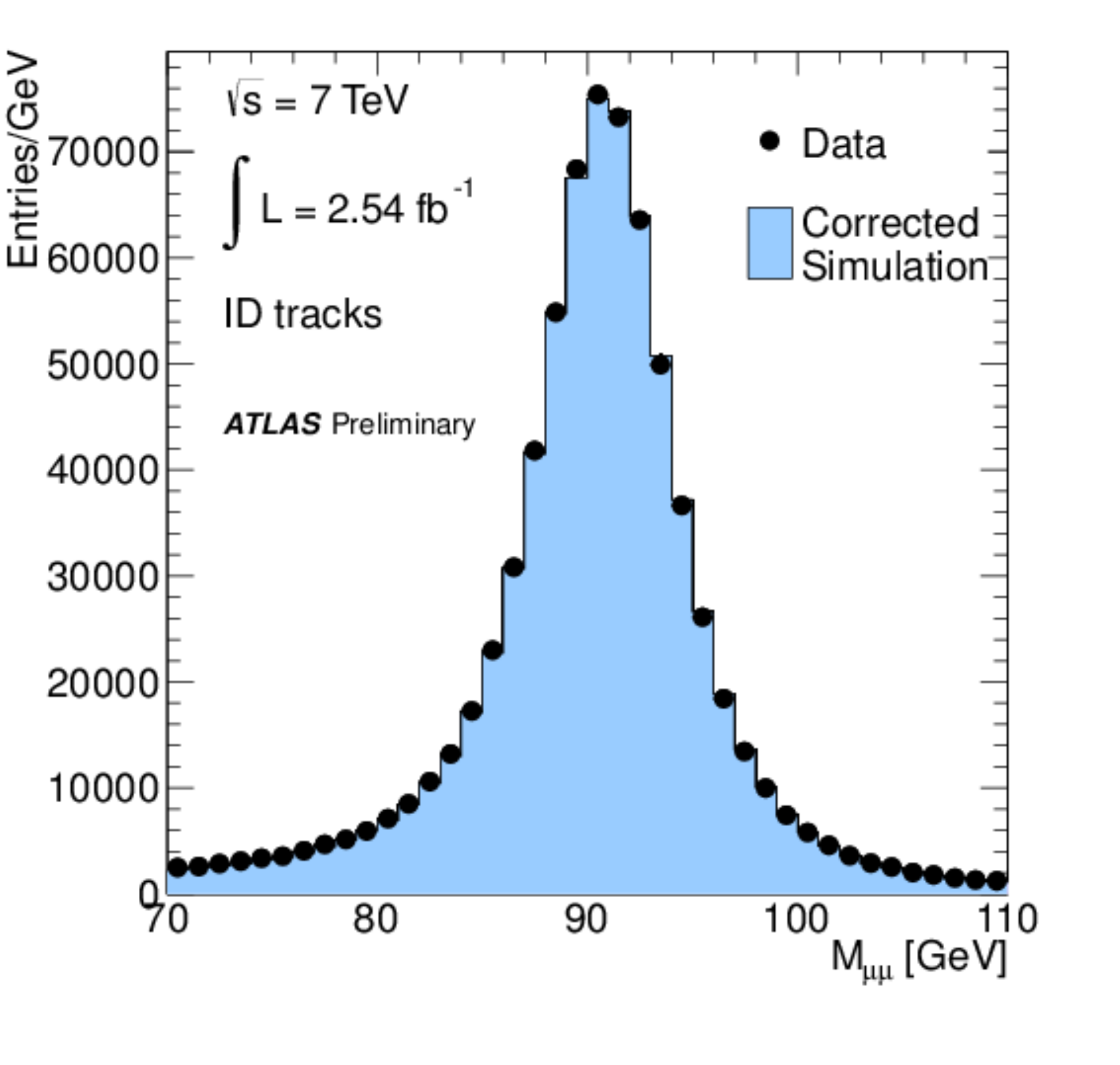}}
  \resizebox{0.495\columnwidth}{!}{\includegraphics{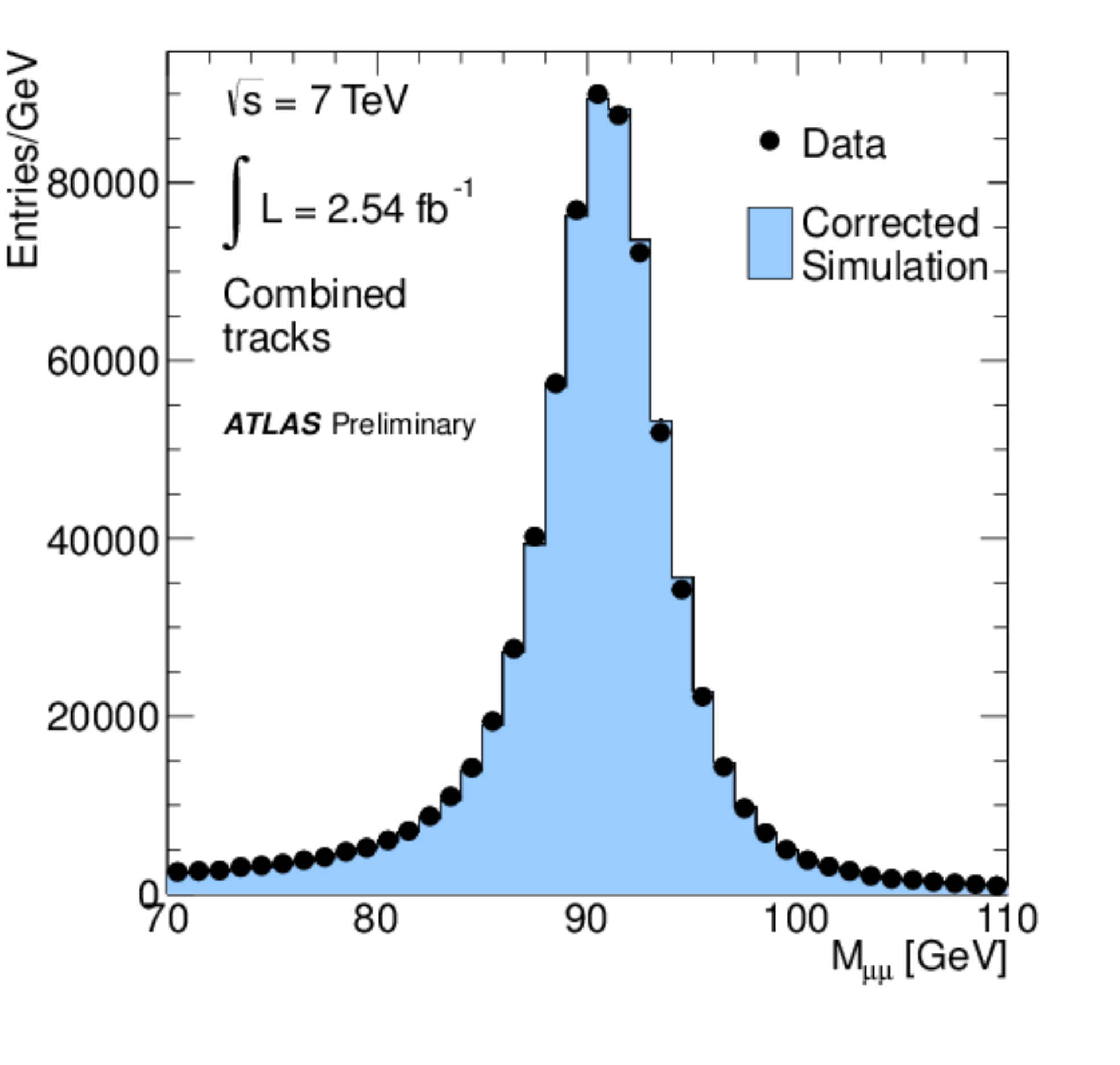}}
  \caption{Di-muon invariant mass comparison in the Z boson mass range between collision data (dots) 
    and simulation (full histogram), after correcting the simulated muon \pT, in the full range of 
    $\eta$. From top left to bottom: MS, ID and combined measureaments are shown.}
  \label{fig:smearcheck}
\end{figure}

The parametrized resolution as a function of \pT for the barrel region, obtained using the 
values of the parameters from the combined fits, are shown in figure \ref{fig:plotMSID} for 
the MS and the ID. The resolution curves for experimental data (in blue) are compared to those
from uncorrected parameters obtained for the simulation (in red).
To indicate the goodness of the simulation correction provided in section \ref{sec:results},
figure \ref{fig:smearcheck} shows the distribution of the di-muon invariant mass in the Z
region after applying the corrections, for MS, ID and combined tracks.

\section{Conclusions}
\label{sec:concl}
A determination of the muon momentum resolution is presented for the integrated luminosity
of 2.54 \fb collision data collected in 2011 with the ATLAS detector. \Zmumu decays have been 
used to evaluate the resolution as a function of the muon \pT and $\eta$, for both the MS and 
the ID. The momentum resolutions were measured on the experimental data and compared with the 
simulation. Results obtained present an improvement of the resolution and the alignment with 
respect to those obtained with 2010 data \cite{2010data}.

\end{document}